# Multi-Robot Adversarial Patrolling:
# Facing a Full-Knowledge Opponent

**Noa Agmon**                                          AGMON@CS.UTEXAS.EDU
*Department of Computer Science*
*University of Texas at Austin*
*TX, USA*

**Gal A Kaminka**                                      GALK@CS.BIU.AC.IL
**Sarit Kraus**                                        SARIT@CS.BIU.AC.IL
*Computer Science Department*
*Bar Ilan University*
*Israel*

## Abstract

The problem of adversarial multi-robot patrol has gained interest in recent years, mainly due to its immediate relevance to various security applications. In this problem, robots are required to repeatedly visit a target area in a way that maximizes their chances of detecting an adversary trying to penetrate through the patrol path. When facing a strong adversary that knows the patrol strategy of the robots, if the robots use a *deterministic* patrol algorithm, then in many cases it is easy for the adversary to penetrate undetected (in fact, in some of those cases the adversary can guarantee penetration). Therefore this paper presents a *non-deterministic* patrol framework for the robots. Assuming that the strong adversary will take advantage of its knowledge and try to penetrate through the patrol's weakest spot, hence an optimal algorithm is one that maximizes the chances of detection in that point. We therefore present a polynomial-time algorithm for determining an *optimal patrol* under the Markovian strategy assumption for the robots, such that the probability of detecting the adversary in the patrol's weakest spot is maximized. We build upon this framework and describe an optimal patrol strategy for several robotic models based on their movement abilities (directed or undirected) and sensing abilities (perfect or imperfect), and in different environment models - either patrol around a perimeter (closed polygon) or an open fence (open polyline).

## 1. Introduction

The problem of multi-robot patrol has gained interest in recent years (e.g., Ahmadi & Stone, 2006; Chevaleyre, 2004; Elmaliach, Agmon, & Kaminka, 2007; Paruchuri, Tambe, Ordonez, & Kraus, 2007; Amigoni, Gatti, & Ippedico, 2008), mainly due to its immediate relevance to various security applications. In the multi-robot patrol problem, robots are required to repeatedly visit a target area in order to monitor it. Many researchers have focused on a frequency-based approach, guaranteeing that some point-visit frequency criteria are met by the patrol algorithm, for example maximizing the minimal frequency or guaranteeing uniform frequency (e.g., refer to Elmaliach et al., 2007; Chevaleyre, 2004; Almeida, Ramalho, Santana, Tedesco, Menezes, Corruble, & Chevaleyr, 2004).

In contrast, we advocate an approach in which the robots patrol in adversarial settings, where their goal is to patrol in a way that maximizes their chances of detecting an adversary trying to penetrate through the patrol path. Thus the decisions of the adversary must be taken into account. Our objective is, therefore, to develop patrol paths for the robots, such that following these paths the robots will maximize the chance of adversarial detection. The problem of adversarial planning and specifically adversarial patrolling is a wide problem, where generally no computational tractable results exist. This paper presents the problem in a restrictive environment of perimeter patrol by a set of homogenous robots, providing a computational tractable optimal result.

As opposed to frequency-driven approaches, in adversarial settings the point-visit frequency criteria becomes less relevant. Consider the following scenario. We are given a cyclic fence of a length of 100 meters and 4 robots must patrol around the fence while moving at a velocity of $1m/sec$. Clearly, the optimal possible *frequency* at each point around the fence, in terms of maximizing the minimal frequency, is $1/25$, i.e., each location is visited once every 25 seconds. This optimal frequency is achieved if the robots are placed uniformly along the fence (facing the same direction) and move forward without turning around. Assume that it takes an adversary 20 seconds to penetrate the area through the fence. As the robots move in a deterministic path, an adversary knowing the patrol algorithm can guarantee penetration if it simply enters through a position that was recently visited by a patrolling robot. On the other hand, if the robots move non-deterministically, i.e., they turn around from time to time with some probability greater than 0, then the choice of penetration position becomes less trivial. Moreover, if we assume that an adversary may penetrate at any time, it motivates the use of nondeterministic patrol behavior indefinitely.

We first consider the problem of patrolling around a closed polygon, i.e., a perimeter. We introduce a non-deterministic framework for patrol under a first order Markovian assumption for the robots' strategy, in which the robots choose their next move at random with some probability $p$. This probability value $p$ characterizes the patrol algorithm. We model the system as a Markov chain (e.g., Stewart, 1994), and using this model we calculate in polynomial time the probability of penetration detection at each point along the perimeter as a function of $p$, i.e., it depends on the choice of patrol algorithm.

Based on the functions defining the probability of penetration detection, we find an optimal patrol algorithm for the robots in the presence of a strong adversary, i.e., an adversary having *full knowledge* on the patrolling robots—their algorithm and current placement. In this case, the adversary uses this knowledge in order to maximize its chances of penetrating without being detected. It is therefore assumed that the adversary will penetrate through the weakest spot of the path, hence the goal of the robots is to maximize the probability of penetration detection in that weakest spot. We provide a *polynomial time* algorithm (polynomial in the input size, depending on the number of robots and the characteristics of the environment) for finding an *optimal* patrol for the robots facing this full knowledge adversary. We show that a non-deterministic patrol algorithm is advantageous, and guarantees some lower bound criteria on the performance of the robots, i.e., on their ability to block the adversary.

We then use the patrol framework to consider additional environment and robotic models. Specifically, we consider the case in which the robots are required to patrol along an open polyline (fence). We show that although this case is inherently different from patrolling along a perimeter, the basic framework can still be used (with some changes) in order to find an optimal patrol algorithm for the robots. We investigate also different movement models of robots, namely the robots can have directionality associated with their movement (and turning around could cost the system time), or they can be omnidirectional. In addition, we model various types of sensing capabilities of the robots, specifically, their sensing capabilities can be perfect or imperfect, local or long-range. In all these cases we show how the basic framework can be extended to include the various models.

This paper is organized as follows. In Section 2 we discuss previous work, related to our research. Section 3 describes the basic robot and environment model. We introduce in Section 4 a framework for the patrolling robots, and describe a polynomial-time algorithm for determining the probability of penetration detection at every point along the patrol path (Section 4.2). We then show in Section 4.3 an algorithm for defining an *optimal* patrol algorithm for the robots, assuming they face a full-knowledge adversary, and in Section 4.4 provides some interesting results from an implementation of the algorithms. In Section 5 we show how the basic framework can be used in order to consider various robotic sensing and movement models and in a different environment (open fences). Section 6 concludes.

## 2. Related Work

Systems comprising multiple robots that cooperate to patrol in some designated area have been studied in various contexts (e.g., Chevaleyre, 2004; Elmaliach, Agmon, & Kaminka, 2009). Theoretical (e.g., Chevaleyre, 2004; Elmaliach et al., 2009; Amigoni et al., 2008) and empirical (e.g., see Sak, Wainer, & Goldenstein, 2008; Almeida et al., 2004) solutions have been proposed in order to assure quality patrol. The definition of quality depends on the context. Most studies concentrate on the *frequency-based patrolling*, which optimizes frequency of visits throughout the designated area (e.g. refer to Elmaliach et al., 2009; Almeida et al., 2004; Chevaleyre, 2004). Efficient patrol, in this case, is a patrol guaranteeing a high frequency of visits in all parts of the area. In contrast, *adversarial patrolling* (addressed in this paper) deals with the detection of moving adversaries who are attempting to avoid detection. Here, an efficient patrol is one that deals efficiently with intruders (e.g., see Sak et al., 2008; Basilico, Gatti, & Amigoni, 2009b; Amigoni et al., 2008).

The first theoretical analysis of the frequency-based multi-robot patrol problem that concentrated on frequency optimization was presented by Chevaleyre (2004). He introduced the notion of *idleness*, which is the duration each point in the patrolled area is not visited. In his work, he analyzed two types of multi-robot patrol schemes on graphs with respect to the idleness criteria: partitioning the area into subsections, where each section is visited continuously by one robot; and the cyclic scheme in which a patrol path is provided along the entire area and all robots visit all parts of the area, consecutively. He proved that in the latter approach, the frequency of visiting points in the area is considerably higher. Almeida et al. (2004) offered an empirical comparison between different approaches towards patrolling with regards to the idleness criteria, and shows great advantage of the cycle based approach.

Elmaliach et al. (2007, 2009) offered new frequency optimization criteria for evaluating patrol algorithms. They provide an algorithm for multi-robot patrol in continuous areas that is proven to have maximal minimal frequency as well as uniform frequency, i.e., each point in the area is visited with the same highest-possible frequency. Their work is based on creating one patrol cycle that visits all points in the area in minimal time, and the robots simply travel equidistantly along this patrol path.

Sak et al. (2008) considered the case of multi-agent adversarial patrol in general graphs (rather than perimeters, as in our work). They concentrated on an empirical evaluation (using a simulation) of several non-deterministic patrol algorithms that can be roughly divided into two: Those that divide the graph between the patrolling agents, and those that allow all agents to visit all parts of the graph. They considered three types of adversaries: random adversary, an adversary that always chooses to penetrate through a recently-visited node and an adversary that uses statistical methods to predict the chances that a node will be visited soon. They concluded that there is no patrol method that outperformed the others in all the domains they have checked, but the optimality depends on the graph structure. In contrast to this investigation, we provide theoretical proofs of optimality for different settings.

The work of adversarial multi-robot patrol was examined also by using game-theoretic approaches (e.g., see Basilico et al., 2009b; Basilico, Gatti, & Amigoni, 2009a; Pita, Jain, Ordonez, Tambe, Kraus, & Magorii-Cohen, 2009; Paruchuri, Tambe et al., 2007). Note that the work described herein can be modeled as a game theoretic problem: Given two players, the robots and the adversary, with a possible set of actions by each side, determine the optimal policy of the robots that will maximize their utility gained from adversarial detection. This is a zero-sum game. Since we assume a strong (full knowledge) adversarial model, we adopt the *minmax* approach, namely, minimizing the maximal utility of the opponent (or in this case: equivalent to maximizing the minimal probability of detection of the robots). However, in our work we do not use game theoretic tools for finding the equilibrium strategy, but use tailored ad-hoc solution that finds the optimal policy for the robots in polynomial time, taking into account the robots' possible sensing and movement capabilities.

The most closely related work by Amigoni et al. (2008) and Basilico et al. (2009b, 2009a) used a game-theoretic approach using *leader-follower games* for determining the optimal strategy for a single patrolling agent. They considered an environment in which a patrolling robot can move between any two nodes in a graph, as opposed to the perimeter model we use. Their solution is suitable for one robot in heterogenous environments, i.e., the utility of the agent and the adversary changes along the vertices of the graph. They formulate the problem as a mathematical programming problem (either multilinear programming or mixed integer linear programming). Consequently, the computation of the optimal strategy is exponential, yet using optimization tools they manage to get good approximation to the optimal solution.

Paruchuri, Tambe et al. (2007) considered the problem of placing *stationary* security checkpoints in adversarial environments. Similar to our assumptions, they assume that their agents work in an adversarial environment in which the adversary can exploit any predictable behavior of the agents, and that the adversary has *full knowledge* of the patrolling agents. They model their system using *Stackelberg games*, which uses policy randomization in the agents' behavior in order to maximize their rewards. The problem is formulated as a linear program for a single agent, yielding an optimal solution for that case. Using this single agent policy, they present a heuristic solution for multiple agents, in which the optimal solution is intractable. Paruchuri, Pearce et al. (2007) further study this problem in cases where the adversarial model is unknown to the agents, although the adversary still has full knowledge of the patrol scheme. They again provide heuristic algorithms for optimal strategy selection by the agents. Pita et al. (2009) continued this research to consider the case in which the adversaries make their choices based on their bounded rationality or uncertainty, rather than make the optimal game-theoretic choice. They considered three different types of uncertainty over the adversary's choices, and provided new mixed-integer linear programs for Stackelberg games that deal with these types of uncertainties.

As opposed to all these works that are based on using game-theoretic approaches and provide approximate or heuristic solutions to intractable optimal solutions, in our work we focus on specific characteristics of the robots and the environment, and provide *optimal polynomial-time* algorithms for finding an optimal patrol strategy for the *multi*-robot team using the minmax approach.

Theoretical work based on stochastic processes that is related to our work is the *cat and mouse* problem (Coppersmith, Doyle, Raghavan, & Snir, 1993), also known as the *predator-prey* (Haynes & Sen, 1995) or *pursuit evasion* problem (Vidal, Shakernia, Kim, Shim, & Sastry, 2002). In this problem, a cat attempts to catch a mouse in a graph where both are mobile. The cat has no knowledge about the mouse's movement, therefore as far as the cat is concerned, the mouse travels similarly to a simple random walk on the graph. We, on the other hand, have worst case assumptions about the adversary. We consider a *robotic* model, in which the movement of the cat is correlated with the movement of a robot, with possible directionality of movement, possible cost of changing directions and possible sensorial abilities. Moreover, in our model the robots travel around a perimeter or a fence, rather than in a general graph. Thus in a sense, our research is concerned with pursuit-evasion on a polyline - open or closed.

Other theoretical work by Shieh and Calvert (1992), based on computational geometry solutions, attempts to find optimal viewpoints for patrolling robots. They try to maximize the view of the robots in the area, show that the problem is $\mathcal{NP}$-Hard, and find approximation algorithms for the problem.

## 3. Robot and Environment Model

In the following section, we provide a description of the robotic model, environment model and the adversarial model. We describe the basic model of patrolling around a perimeter (closed polygon). Further environments and robotic models are discussed in Section 5.

### 3.1 The Environment

We consider a patrol in a circular path around a closed polygon $P$. The path around $P$ is divided into $N$ segments of a length of uniform *time distance*, i.e., each robot travels through one segment per cycle while sensing it (its velocity is 1 segment / 1 time cycle). This division into segments makes it possible to consider patrols in heterogeneous paths. In such areas, the difficulty of passing through terrains varies from one terrain to another, for example driving in muddy tracks vs. driving on a road. In addition, riding around corners requires a vehicle to slow down. Figure 1 demonstrates a transition from a given area to a discrete cycle. The area, on the left, is given along with its velocity constraints. The path is then divided into segments such that a robot travels through one segment per time cycle *while monitoring it*, i.e., the length of each segment is determined by both the velocity of the robot (corresponding to the time it takes it to travel through the specific segment) and the sensorial capabilities of the robot. After the path is divided into segments with uniform travel time, it is equivalent to considering a simple cycle as appears in the right of Figure 1.

Note that the distance between the robots is calculated with respect to the number of segments between them, i.e., the distance is in travel time. For example, if we say that the distance between $R_1$ and $R_2$ is 7, then there are 7 segments between them, and if $R_1$ had remained still, then it would have taken $R_2$ 7 time cycles to reach $R_1$ (assuming $R_2$ is headed towards the right direction).

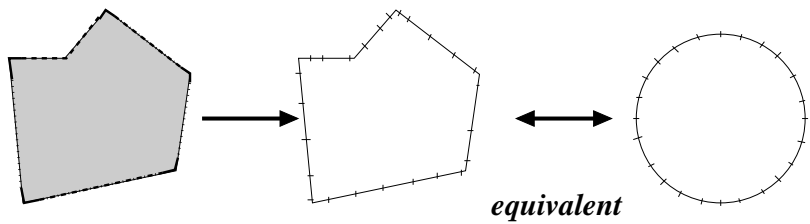

*equivalent*

**Figure 1**: An example for creating discrete segments from a circular path with the property that the robots travel through one segment per cycle. The different line structures along the perimeter on the left correlate to different velocity constraints, which are converted (in the middle figure) to $N$ segments in which the robots travel during one time cycle. This figure is equivalent to the figure in the right, which is a simple cycle divided into $N$ uni-time segments.

## 3.2 Patrolling Robotic Model

We consider a system of $k > 1$ homogenous mobile robots $R_1, \ldots, R_k$, that are required to patrol around a closed polygon. The robots operate in cycles, where each cycle consists of two stages.

1. **Compute**: Execute the given algorithm, resulting in a goal point, denoted by $p_G$, to which the robot should travel.

2. **Move**: Move towards point $p_G$.

This model is synchronous, i.e. all robots execute each cycle simultaneously. We concentrate our attention on the *Compute* stage, i.e., how to compute the next goal point.

We assume the robots' movement model is directed such that if $p_G$ is behind the robot, it has to physically turn around. Turning around is a costly operation, and we model this cost in time, i.e., if the robot turns around it resides in its segment for $\tau$ time units. The case in which the movement model is not directed is discussed in Section 5.1. Throughout the paper we assume for simplicity that $\tau = 1$, unless stated otherwise.

A key result of this research (Section 4) is that optimal patrolling necessitates robots to be placed at a uniform distance $d = N/k$ from one another along the perimeter. Consequently, we require the robots to be coordinated in the sense that all robots move in the same direction, and if decided to turn around they do it simultaneously. This requirement guarantees that the uniform distance of $d$ is maintained throughout the execution of the patrol algorithm. Note that this tightly-coordinated behavior is achievable in centralized systems, or in systems where communication exists between all team members. Other practical implementations may exist (for example uniformly seeding a pseudorandom number generator for all the robots), but they all require coordination inside the team. Distributed systems that cannot assume reliable communication are left for future work.

### 3.3 Adversarial Model

Our basic assumption is that the system consists of an adversary that tries to penetrate once through the patrolling robots' path without being detected. The adversary decides, at any unknown time, through which segment to penetrate. Its penetration time is *not instantaneous*, and lasts $t$ time units, during which it stays at the same segment.

**Definition 1.** *Let $s_i$ be a discrete segment of a perimeter $P$ which is patrolled by one robot or more. Then the* Probability of Penetration Detection *in $s_i$, $\mathsf{ppd}_i$, is the probability that a penetrator going through $s_i$ during $t$ time units will be detected by some robot going through $s_i$ during that period of time.*

In other words, $\mathsf{ppd}_i$ is the probability that a patrol path of some robot will pass through segment $s_i$ during the time that a penetration is attempted through that segment, hence it is calculated for each segment with respect to the current location of the robots at a given time (since the robots maintain uniform distance between them throughout the execution, this relative location remains the same at all times). We use the general acronym $\mathsf{ppd}$ when referring to the general term of probability of penetration detection (without reference to a certain segment).

Recall that the time distance between every two consecutive robots around the perimeter is $d = N/k$. Therefore we consider $t$ values between the boundaries $\frac{d+\tau}{2} \leq t < d$. The reason for this is that if it takes the robot $\tau$ time units to turn, then the robot adjacent to $s_0$ will have probability $> 0$ of arriving at every segment $s_i, 0 \leq i \leq t$, while the robot adjacent to $s_d$ has probability $> 0$ of arriving at segments $s_i, d - (t - \tau) \leq i \ leqd$. Hence segment $s_{t+1}$ has probability $> 0$ of being visited only if $d - (t - \tau) \leq t + 1 \Rightarrow \frac{d+\tau+1}{2} \leq t$, otherwise there is at least one segment, $s_{t+1}$, that has probability 0 of being visited during $t$ time units. Therefore an adversary having full knowledge on the patrol will *always* manage to successfully penetrate regardless of the actions taken by the patrolling robots. Note that $\tau$ appears in this equation since it influences the number of segments reachable by the robot located in segment $s_{d+1}$ if turning around $(s_d, s_{d-1}, \ldots, s_{d/2+\tau})$. On the other hand, if $t \geq d$ then *all* segments $s_i$ can have $\mathsf{ppd}_i = 1$ simply by using a deterministic algorithm.

We define the *patrol scheme* of the robots as the

1. Number of robots, the distance between them and their current position.

2. The movement model of the robots and any characterization of their movement.

3. The robots' patrol algorithm.

The patrol scheme reflects the knowledge obtained by the adversary on the patrolling robots at any given time (hence is not necessarily time dependent).

We consider a strong adversarial model in which the adversary has *full knowledge* of the patrolling robots. Therefore the full knowledge adversary knows the patrol scheme, and will take advantage of this knowledge in order to choose its penetration spot as the *weakest spot* of the patrol, i.e., the segment with minimal ppd. The solution concept adopted here (as stated in Section 2) is similar to the game-theoretic *minmax* strategy, yielding a strategy that is in equilibrium (none of the players—robots or adversary—has any initiative to diverge from their strategy). The adversary can learn the patrol scheme by observing the behavior of the robots for a sufficient amount of time. Note that in security applications, such strong adversaries exist. In other applications, the adversary models the behavior of the system in the "worst case scenario" from the patrolling robots point of view (similar to the classical Byzantine fault model in distributed systems, see Lynch, 1996).

In our environment, the robots are responsible only for *detecting* penetrations and *not* handling the penetration (which requires task-allocation methods). Therefore the case in which the adversary issues multiple penetrations is similar to handling a single penetration, as the robots detect, report and continue to monitor the rest of the path, according to their algorithm.

## 4. A Framework for Adversarial Patrolling of Perimeters

The environment we consider is a linear environment, in which at each step the robots can decide to either go straight or turn around. The framework we suggest is *nondeterministic* in the sense that at each time step the decision it done independently, at random, with some probability $p$. Formally,

$$\text{Probability of next move} = \begin{cases} p & \text{Go straight} \\ 1-p & \text{Turn around} \end{cases}$$

Since the different patrol algorithms we consider vary in the probability $p$ of the next step, we assert that the probability $p$ *characterizes the patrol algorithm*.

Assume a robot is currently located in segment $s_i$. Therefore if the robot is facing segment $s_{i+1}$, then with a probability of $p$ it will go straight to it and with a probability of $1-p$ it will turn around and face segment $s_{i-1}$. Similarly, if it is facing segment $s_{i-1}$, then with a probability of $p$ it will reach segment $s_{i-1}$ and with a probability of $1-p$ it will face segment $s_{i+1}$.

Note that the probability of penetration detection in each segment $s_i$, $1 \leq i \leq d$, is determined by probability $p$ characterizing the patrol algorithm, therefore $ppd_i$ is a function of $p$, i.e., $ppd_i(p)$. However, whenever possible we will use the abbreviation $ppd_i$. By the definition of $ppd_i$, we need to find the probability that $s_i$ will be visited during $t$ time units by some robot. Assuming perfect detection capabilities of the robots, $ppd_i$ is determined only by the *first* visit of some robot to $s_i$, since once the intruder is detected the detection mission is successful (specifically, once the segment is visited, the "game" is over). Note that $ppd_i$ is calculated *regardless of the actions of the adversary*.

As stated previously, in order to guarantee optimality of the patrol algorithm, the robots should be uniformly distributed along the perimeter with a distance of $d = N/k$ between every two consecutive robots, and that they are coordinated in the sense that if they are supposed to turn around, they do so simultaneously. In the following theorem and supporting lemmas we prove optimality of these assumptions in a full-knowledge adversarial environment.

Lemma 1 follows directly from the fact that the movement of the robots is continuous, thus a robot $R_l$ cannot move from segment $s_i$ to segment $s_{i+j}$, $j > 0$, without visiting segments $s_{i+1}, \ldots, s_{j-1}$ in between. Note that since $k > 1$ it follows that the number of segments unvisited by $R_l$ is greater than $2t$ (otherwise a simple deterministic algorithm would suffice to detect the adversary with probability 1). Therefore during $t$ time units $R_l$ residing initially in segment $s_0$ cannot visit segment $s_i$, $i < t$, arriving from the other direction of the perimeter without visiting the segments closer to its current location ($s_0$) first (this argument holds for segments to the left and to the right of $s_0$).

**Lemma 1.** *For a given $p$, the function $\mathsf{ppd}_i^l : \mathbb{N} \Rightarrow [0,1]$ for constant $t$ and $R_l$ residing in segment $s_0$ is a monotonic decreasing function, i.e., as the distance between a robot and a segment increases, the probability of reaching it during $t$ time units decreases.*

**Lemma 2.** *As the distance between two consecutive robots along a cyclic patrol path is smaller, the $\mathsf{ppd}$ in each segment is higher and vice versa.*

*Proof.* Consider a sequence $S_1$ of segments $s_1, \ldots, s_w$ between two adjacent robots, $R_l$ and $R_r$, where $s_1$ is adjacent to the current location of $R_l$ and $s_w$ is adjacent to the current location of $R_r$. Let $S_2$ be a similar sequence, but with $w - 1$ segments, i.e., the distance between $R_l$ and $R_r$ decreases by one segment. Assume that other robots are at a distance greater than or equal to $w - 1$ from $R_l$ and $R_r$, and that $w - 1 < t$. Since a robot may influence the $\mathsf{ppd}$ in segments that are up to a distance $t$ from it (as it has a probability of 0 of arriving at any segment at a greater distance within $t$ time units), the probability of penetration detection, $\mathsf{ppd}$, in these sequences is influenced only by the possible visits of $R_l$ and $R_r$.

Denote the probability of penetration detection in segment $s_i \in S_j$ by $\mathsf{ppd}_i(j)$, $1 \leq i \leq w$, $j \in \{1, 2\}$, and the probability that the penetrator will be detected by robot $R_x$ by $\mathsf{ppd}_i^x(j)$, $x \in \{l, r\}$. Therefore, for any segment $s_i \in S_j$, $\mathsf{ppd}_i(j) = \mathsf{ppd}_i^l(j) + \mathsf{ppd}_i^r(j) - \mathsf{ppd}_i^l(j)\mathsf{ppd}_i^r(j)$ (either $R_l$ or $R_r$ will detect the adversary, not both). Note that either $\mathsf{ppd}_i^l(j)$, $\mathsf{ppd}_i^r(j)$ or both can be equal to 0. We need to show that $\mathsf{ppd}_i(2) \geq \mathsf{ppd}_i(1)$, for all $1 \leq i \leq w$, and for at least one segment $s_m$, $\mathsf{ppd}_m(2) > \mathsf{ppd}_m(1)$. Specifically, it is sufficient to show that $\mathsf{ppd}_i^l(2) + \mathsf{ppd}_i^r(2) - \mathsf{ppd}_i^l(2)\mathsf{ppd}_i^r(2) - \{\mathsf{ppd}_i^l(1) + \mathsf{ppd}_i^r(1) - \mathsf{ppd}_i^l(2)\mathsf{ppd}_i^r(2)\} \geq 0$, and for some $i$ this inequality is strict.

For every segment $s_i$, $\mathsf{ppd}_i^l(1) = \mathsf{ppd}_i^l(2)$ (there is no change in its relative location), hence we need to prove that $\mathsf{ppd}_i^r(2) - \mathsf{ppd}_i^r(1) \geq \mathsf{ppd}_i^l(2)\{\mathsf{ppd}_i^r(2) - \mathsf{ppd}_i^r(1)\}$. Since $0 \leq \mathsf{ppd}_i^l(2) \leq 1$, in order for the inequality to hold, it is left to show that $\mathsf{ppd}_i^r(2) - \mathsf{ppd}_i^r(1) \geq 0$. From Lemma 1 we know that $\mathsf{ppd}_i^r(j)$ is monotonically decreasing, therefore for each $i$, $\mathsf{ppd}_i^r(2) \geq \mathsf{ppd}_i^r(1)$, which completes the proof of this inequality.

It is left to show that for some $i = m$, $\mathsf{ppd}_m^r(2) - \mathsf{ppd}_m^r(1) > \mathsf{ppd}_m^l(2)\{\mathsf{ppd}_m^r(2) - \mathsf{ppd}_m^r(1)\}$, i.e., for some $m$ in which $\mathsf{ppd}_m^l(2) \neq 1$, $\mathsf{ppd}_m^r(2) > \mathsf{ppd}_m^r(1)$. Robot $R_r$ may influence the $\mathsf{ppd}$ on both of his sides - segments located to the left and to the right of its current position. Denote the number of influenced segments to its right by $y$ ($y$ may be equal to 0). If $y > 0$, then $\mathsf{ppd}_{w-y+1}^r(2) > \mathsf{ppd}_{w-y}^r(1)$. In other words, $R_r$ has a probability of 0 of reaching the segment with a distance of $t + 1$ from it in $S_1$, but in $S_2$ it is $y$ segments away from it, therefore $R_r$ has a probability greater than 0 to reach it. If $y = 0$, then $\mathsf{ppd}_w^r(2) = 1 > \mathsf{ppd}_w^r(1)$, as $R_r$ lies exactly in segment $s_w$ of $S_2$, and $\mathsf{ppd}_w^r(1) = 0$. □

**Theorem 3.** *A team of $k$ mobile robots engaged in a patrol mission maximizes minimal* $\mathsf{ppd}$ *if the following conditions are satisfied.* **a.** *The time distance between every two consecutive robots is equal* **b.** *The robots move in the same direction and speed.*

Note that condition **b** means that all robots move together in the same direction, i.e., if they change direction, then all $k$ robots change their direction simultaneously.

*Proof.* Following Lemma 2, it is sufficient to show that the combination of conditions **a** and **b** yield the minimal distance between two consecutive robots along the cyclic path. Since we have $N$ segments and $k$ robots, there are $\binom{N}{k}$ possibilities of initial placement of robots along the cycle (robots are homogenous, so this is regardless of their order). If the robots are positioned uniformly along the cycle, then the time distance between each pair of consecutive robots is $N/k$. This is the minimal value that can be reached. Therefore, clearly, condition **a** guarantees this minimality.

If the robots are not coordinated, then it is possible that two consecutive robots along the cycle, $R_i$ and $R_{i+1}$, will move in opposite directions. Therefore the distance between them will increase from $\frac{N}{k}$ to $\frac{N}{k} + 2$, and by Lemma 2 the $\mathsf{ppd}$ in the segments between them will be smaller. If $R_i$ and $R_{i+1}$ move towards one another, then the distance between them will be $\frac{N}{k} - 2$ and the $\mathsf{ppd}$ in the segments between them will become higher. On the other hand, some pair $R_j$ and $R_{j+1}$ exists such that the distance between them increases, as the total sum of distances between consecutive robots remains $N$, hence the minimal $\mathsf{ppd}$ around the cycle will become smaller.

Therefore the only way of achieving the minimal distance (maximal $\mathsf{ppd}$) is by assuring that condition **a** is satisfied, and maintaining it is achieved by satisfying condition **b**. □

Since when facing a full-knowledge adversary, the goal of the robots is to maximize the minimal $\mathsf{ppd}$ along the perimeter, the following corollary follows.

**Corollary 4.** *In the full-knowledge adversarial model, an optimal patrol algorithm must guarantee that the robots are positioned uniformly along the perimeter throughout the execution of the patrol.*

## 4.1 The Penetration Detection Problem

The general definition of the problem is as follows.

**Penetration detection (*PD*) problem:** Given a circular fence (perimeter) that is divided into $N$ segments, $k$ robots uniformly distributed around this perimeter with a distance of $d = N/k$ (in time) between every two consecutive robots, assume that it takes $t$ time units for the adversary to penetrate, and the adversary is known to have full-knowledge of the patrol scheme. Let $p$ be the probability characterizing the patrol algorithm of the robots, and let $\mathsf{ppd}_i(p)$, $1 \leq i \leq d$ be a description of $\mathsf{ppd}_i$ as a function of $p$. Find the optimal value $p$, $p_{opt}$, such that the minimal ppd throughout the perimeter is maximized. Formally,

$$p_{opt} = \underset{0 \leq p \leq 1}{\arg\max} \{ \min_{1 \leq i \leq d} \mathsf{ppd}_i(p) \}$$

To summarize the model and the Theorems presented above, an optimal algorithm for multi-robot perimeter patrol under the Markovian strategy assumption for the robots has the following characteristics.

---

- The robots are placed uniformly around the perimeter with $d$ segments between every two consecutive robots.

- The robots are coordinated in the sense that if they decide to turn around, then they do it simultaneously.

- At each time step, the robots continue straight with a probability of $p$ or turn around with a probability of $1-p$, and if they turn around they stay in the same segment for $\tau$ time units.

---

Note that under the above framework (i.e. the framework for homogenous robots), the division of the perimeter into sections of $d$ segments creates an equivalent symmetric environment in the sense that in order to calculate the optimal patrol algorithm it is sufficient to consider only *one section* of $d$ segments, and not the entire perimeter of $N$ segments. This is due to the fact that each section is completely equivalent to the other, and remains so throughout the execution.

We divide the goal of solving the *PD* problem, i.e., finding an *optimal* patrol algorithm into two stages.

1. Calculating the $d$ $\mathsf{ppd}_i$ functions for each $1 \leq i \leq d$. This is determined according to the robotic movement model (directed or undirected), environment model (perimeter/fence) and sensorial model (perfect/imperfect, local/extended).

2. Given the $d$ $\mathsf{ppd}_i$ functions, find the solution to the *PD* problem, i.e., maximize the ppd in the segment(s) with minimal ppd.

These two steps are independent in the sense that incorporating various different robotic models will not change the process of determining the solution to the *PD* problem, as long as the result of the procedure are $d$ functions representing the ppd values in each segment.

On the other hand, if we would like to consider different goal functions other than maximizing the minimal ppd (for example maximizing the *expected* ppd), it can be done without any change in the first stage, i.e., determining the ppd functions. The important result is that this framework can be applied to both different environment and robotic models (for example fence patrol), *and* different goal functions (corresponding to different adversarial settings).

The first stage for a the basic model (perimeter patrol, directed movement model of the robots, robots with perfect local sensing) is described in Section 4.2, and the second stage is described in Section 4.3. Extensions of the first stage to different robot motion models and sensing models are described in Section 5.

## 4.2 Determining the Probability of Penetration Detection

In order to find an *optimal* patrol algorithm, it is necessary to first determine the probability of penetration detection at each segment $s_i$ ($\mathsf{ppd}_i$), which is a function of $p$ (the probability characterizing the patrol algorithm, as shown in Section 4.1). In this section we present a polynomial time algorithm that determines this probability.

As stated previously, based on the symmetric nature of the system, we need to consider only one section of $d$ segments that lie between two consecutive robots, without loss of generality, $R_1$ and $R_2$. We use a Markov chain in order to model the possible states and transition between states in the system.

In order to calculate the probability of detection in each segment along $t$ time cycles, we use the graphic model $G$ illustrated in Figure 2. For each segment $s_i$ in the original path, $1 \leq i \leq d$, we create two states in $G$: One for moving in a clockwise direction ($s_i^{cw}$), and the other for moving in a counterclockwise direction ($s_i^{cc}$). If $R_1$ or $R_2$ reach one of the $s_i$ segments within $t$ time units, then the adversary is discovered, i.e., it does not matter if the segment is visited more than once during these $t$ time units. Therefore we would like to calculate only the probability of the *first* arrival to each segment, and this is done by defining the state $s_{dt}$ (corresponding to $s_0$ and $s_0'$) as *absorbing states*, i.e., once a robot passes through $s_i$ once, its additional visits to this segment in this path will not be considered. The edges of $G$ are as follows. One outgoing edge from $s_i^{cw}$ to $s_i^{cc}$ exists with a probability of $1-p$ for turning around, and one outgoing edge to $s_{i-1}^{cw}$ exists with a probability of $p$ for continuing straightforward. Similarly, one outgoing edge from $s_i^{cc}$ to $s_i^{cw}$ exists with a probability $1-p$ for turning around, and one outgoing edge to $s_{i+1}^{cc}$ exists with a probability of $p$ for continuing straightforward.

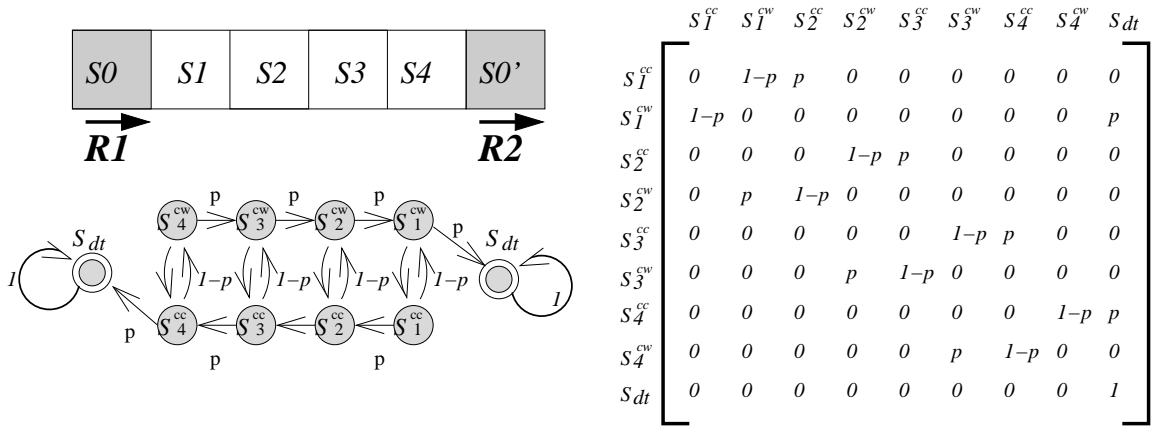

**Figure 2**: Conversion of the initial segments and robot locations into a graphical model, and the respective stochastic matrix $M$. Each segment corresponds to two states: one going clockwise and one going counterclockwise. $\mathsf{ppd}_i$ are all paths starting from $s_i^{cw}$ and ending at $s_{dt}$.

In the following theorem, we prove that the probability of detecting the adversary by some robot in segment $s_i$ (i.e., the probability of arriving at a segment during $t$ time units) is equivalent to finding all paths of size at most $t$ to the absorbing state starting at state $s_i^{cc}$. Therefore it is possible to use the Markov chain representation for determining $\mathsf{ppd}_i$, as shown in Algorithm FindFunc.

**Theorem 5.** *Determining the probability of penetration detection at segment $s_i$, $\mathsf{ppd}_i$, is equivalent to finding all paths of length at most $t$ that start at $s_i^{cw}$ and end in $s_{dt}$ in the Markov chain described above.*

*Proof.* For simplicity reasons, in this proof we distinguish between $s_{dt}^l$ and $s_{dt}^r$, which are the absorbing state to the left and to the right of the Markov chain (respectively), although practically they are represented by the same state $s_{dt}$.

Clearly, due to the $d$ and $t$ values considered, $\mathsf{ppd}_i$ is determined only by the visits of the two robots surrounding the section of $d$ segments $s_1, \ldots, s_d$, denoted by $R_l$ and $R_r$. Recall that the probability of penetration detection in segment $s_i$ is defined as $\mathsf{ppd}_i = \mathsf{ppd}_i^l + \mathsf{ppd}_i^r - \mathsf{ppd}_i^l \mathsf{ppd}_i^r$, where $\mathsf{ppd}_i^r$ ($\mathsf{ppd}_i^l$) is the probability that the adversary, penetrating through $s_i$, is detected by $R_r$ ($R_l$). We claim that $\mathsf{ppd}_i^l$ is equivalent to computing the paths starting from $s_i^{cw}$ and ending at the absorbing state $s_{dt}^r$ (similarly $\mathsf{ppd}_i^r$ by state $s_{dt}^l$). Clearly, under this claim, since a path of length at most $t$ cannot reach both $s_{dt}^r$ and $s_{dt}^l$, it follows that $\mathsf{ppd}_i^l \mathsf{ppd}_i^r = 0$, and the theorem will follow. We prove the claim for $\mathsf{ppd}_i^l$, where $\mathsf{ppd}_i^r$ follows directly.

$\mathsf{ppd}_i^l$ is the probability that $R_l$ will reach $s_i$ at least once during $t$ time units. Therefore, we must construct all paths starting from the current location of $R_l$ that passes through $s_i$, but take into account only the first visit to the segment (everything beyond the first visit results anyway in probability of detection $= 1$). At each step $R_l$ continues straight with probability $p$ or turns around with probability $1 - p$. This is equivalent to keeping $R_l$ in place, and moving the segments towards $R_l$ with probability $p$ and switch the segments' direction with probability $1 - p$. Hence, every path starting at state $s_i^{cw}$ (without loss of generality; computing paths starting at $s_i^{cc}$ is equivalent, but requires switching the locations of $R_l$ and $R_r$ in the representation) reaching $s_{dt}^r$ is equivalent to a path started by $R_l$ and passing through $s_i$. Since $s_{dt}^r$ is set to be an absorbing state, every path passing through it will not be considered again, i.e., only the first visit of $R_l$ to $s_i$ is considered, as required. □

Using the Markov chain, we can define the stochastic matrix $M$ which describes the state transitions of the system. Figure 2 illustrates the Markov chain and its corresponding stochastic matrix $M$ used for computing the $\mathsf{ppd}$ functions. The probability of arrival at segment $s_i$ during $t$ time units, hence the probability penetration detection in that segment, is the $s_{2d+1}^{cc} + s_{2d+1}^{cw}$ entry of the result of $V_i \times M^t$, where $V_i$ is a vector of 0's, except for a 1 on the $2i - 1$'th location. The formal description of the algorithm is given by Algorithm 1. Note that the algorithm makes a symbolic calculation, hence the result is a set of $d$ functions of $p$. The time complexity of Algorithm FindFunc depends on the calculation time of $M^t$, which is generally $t \times (2d)^3$. However, since $M$ is sparse, methods for multiplying such matrices efficiently exist (e.g., see Gustavson, 1978), reducing the time complexity to $t(2d)^2$, i.e. $\mathcal{O}(td^2)$. Since $t$ is bounded by $d - 1$, the time complexity is $\mathcal{O}(d^3)$.

---

**Algorithm 1** Algorithm FindFunc($d, t$)

---

1: Create matrix $M$ of size $(2d+1)(2d+1)$, initialized with 0s
2: Fill out all entries in $M$ as follows:
3: $M[2d+1, 2d+1] = 1$
4: **for** $i \leftarrow 1$ to $2d$ **do**
5:    $M[i, \max\{i+1, 2d+1\}] = p$
6:    $M[i, \min\{1, i-2\}] = 1 - p$
7: Compute $MT = M^t$
8: $Res$ = vector of size $d$ initialized with 0s
9: **for** $1 \leq loc \leq d$ **do**
10:    $V$ = vector of size $2d+1$ initialized with 0s.
11:    $V[2loc] \leftarrow 1$
12:    $Res[loc] = V \times MT[2d+1]$
13: Return $Res$

---

### 4.2.1 HANDLING HIGHER VALUES OF $\tau$

Algorithm FindFunc and Figure 2 demonstrate the case in which $\tau = 1$, i.e., if the robot turns around (with probability $1 - p$) it remains in its current position for one time step. In the general case, when the robot turns around, the cost of turning—modeled in time—can be higher. In such cases, the Markov chain is modified to represent the value of $\tau$. Specifically, for each segment $s_i$, instead of having two corresponding states ($s_i^{cw}$ and $s_i^{ccw}$), we have $2(\tau)$ states: $s_i^{cw}$ and $s_i^{ccw}$, and one set of $\tau - 1$ states for turning around to each direction (from $cw$ to $ccw$ and vice versa). The probabilities assigned to each of the edges is $1 - p$ for the first outgoing edge from $s_i^{cc}$ to the first intermediate state towards $s_i^{cw}$ and 1 for each edge on that direction, and similarly on the path from $s_i^{ccw}$ to $s_i^{cc}$. See Figure 3 for an illustration. The matrix $M$ is filled out according to the new chain, and the time complexity of creating this matrix grows in a factor of $\tau$—from $(2d+1)^t$ to $(2\tau d+1)^t$. However, as long as $\tau$ is a constant, the total time complexity does not change.

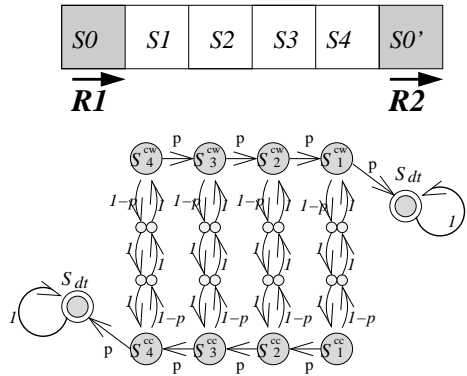

**Figure 3**: Illustration of the Markov chain when $\tau > 1$, and specifically, here $\tau = 3$.

### 4.3 An Optimal Adversarial Patrol Algorithm for Full-Knowledge Adversaries

In cases in which the robots face a full knowledge adversary, it is assumed that the adversary will take advantage of this knowledge to find the weakest spot of the patrol, i.e., the segment with minimal probability of penetration detection. Therefore an optimal patrol algorithm to handle such an adversary is the one that maximizes the minimal ppd throughout the perimeter. Hence we need to find an optimal $p$, $p_{opt}$, such that the minimal ppd throughout the perimeter is maximized.

Also here, since our environment is symmetric, we do not need to consider the entire patrol path, but only a section of $d$ segments between two consecutive robots. The input in this procedure is the set of $d$ $\mathsf{ppd}_i(p)$ functions that were calculated in the previous section (Section 4.2).

After establishing $d$ equations representing the probability of detection in each segment, we must find the $p$ value that maximizes the minimal possible value in each segment, where $p$ is continuous in the range $p \in [0,1]$. Denote these equations by $\mathsf{ppd}_i(p)$, $1 \leq i \leq d$. The maximal minimal value that we are looking is the $p$ value yielding the maximal value inside the intersection of all integrals of $\mathsf{ppd}_i(p)$. The intersection of all integrals is also known as the *lower envelope* of the functions (Sharir & Agarwal, 1996).

Observing the problem geometrically, consider a vertical sweep line that sweeps the section $[0,1]$ and intersects with all $d$ curves. It seeks the point $p$ in which the minimal intersection point between the sweep line and the curves, denoted by $\mathsf{ppd}^*(p)$, is maximal. This $p$ is the maximin point. Since the segment $[0,1]$ and the functions $\mathsf{ppd}_1, \ldots, \mathsf{ppd}_d$ are continuous, this sweep line solution cannot be implemented. We prove in the following lemma that this point is either an intersection point of two curves, or a local maxima of one curve (see Figure 4). See Algorithm 2 for the formal description of Algorithm FindP.

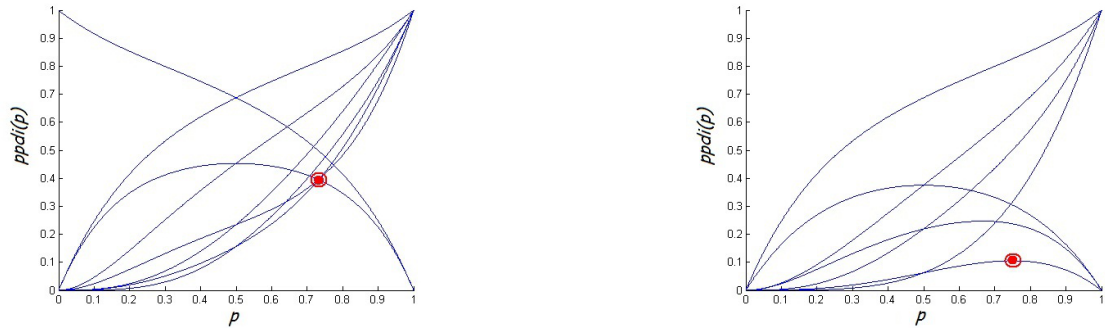

**Figure 4**: An illustration of two possible maximin points (marked by a full circle). The curves represent $d$ $\mathsf{ppd}_i(p)$ functions in $p \in [0,1]$. On the left, the maximin point is created by the intersection of two curves. On in the right, the maximin point it is the local maxima of the lowest curve.

In the following, we prove that Algorithm FindP finds point $p$ such that the maximin property is satisfied.

**Lemma 6.** *A point $p$ yields a maximin value $\mathsf{ppd}^*(p)$ if the following two properties are satisfied.*
**a.** $\mathsf{ppd}^*(p) \leq \mathsf{ppd}_i(p) \ \forall 1 \leq i \leq d.$

**b.** *One of the two following conditions holds:* $\mathsf{ppd}^*(p)$ *is an intersection of two curves (or more),* $\mathsf{ppd}_i(p)$ *and* $\mathsf{ppd}_j(p)$ *or a local maxima of curve* $\mathsf{ppd}_k(p)$.

*Proof.* Property **a.** is derived from the definition of a maximin point. Therefore we are looking for the maximal point that satisfies property **a**. We must still show that this point, $\mathsf{ppd}^*(p)$, is obtained by either an intersection of two or more curves or is a local maxima. Clearly, a maximal point of an integral is found on the border of the integral (the curve itself). The area which is in the intersections of all curves lies beneath parts of curves, $\mathsf{ppd}_{i_1}, \ldots, \mathsf{ppd}_{i_m}$, such that $\mathsf{ppd}_{i_j}$ is the minimal curve in the section between two points $[l^j, r^j]$ and $\bigcup_{j=1}^m [l^j, r^j] = [0, 1]$. By finding the maximal point in each section $\mathsf{ppd}_{max}^j = \max\{f(x), x \in [l^j, r^j]\}$, and choosing the maximal between them, i.e., $\max\{\mathsf{ppd}_{max}^j, 1 \leq j \leq m\}$, we obtain $\mathsf{ppd}^*(p)$. In each section $[l^j, r^j]$ the maximal point can be either inside the section or on the borders of the section. The former case is precisely a local maxima of $\mathsf{ppd}_{i_j}$. The latter is the intersection point of two curves $\mathsf{ppd}_{i_{j-1}}, \mathsf{ppd}_{i_j}$ or $\mathsf{ppd}_{i_j}, \mathsf{ppd}_{i_{j+1}}$. $\square$

**Lemma 7.** *A point $p$ exists yielding a maximin value* $\mathsf{ppd}^*(p) > 0$.

*Proof.* In order to prove the lemma, we need to show that the intersection of all integrals $\mathsf{ppd}_1, \ldots, \mathsf{ppd}_d$ in the $x$ section $[0, 1]$, and the $y$ section $(0, 1]$ is not empty. It suffices to show that for every $\mathsf{ppd}_i$, $\mathsf{ppd}_i(x) > 0, 0 < x < 1$.

Each function $\mathsf{ppd}_i, 1 \leq i \leq d$ represents the $\mathsf{ppd}$ in a segment $s_i$ between two robots. From our requirement that $t \geq \frac{d}{2} + 1$ (for $\tau = 1$), it follows that in all models we consider, for $0 < p < 1$ the $\mathsf{ppd} \neq 0$. Note that if $p = 0$ or $p = 1$, then $\mathsf{ppd}$ is either 0 or 1, but this does not contradict the fact that we have a point guaranteeing $\mathsf{ppd}^*(p) > 0$. $\square$

Algorithm FindP finds this point by scanning all possible points satisfying the conditions given in Lemma 6, and reporting the $x$-value (corresponding to the $p$ value) with a $y$-value dominated by all $\mathsf{ppd}_i$. The input to the algorithm is a vector of functions $\mathsf{ppd}_i$, $1 \leq i \leq d$ and the value $t$. Computing the intersections between every pair of functions costs $d^2 t^2$: $d^2$ for all pair computation, $t^2$ for finding the root of the polynomial using, for example, the Lindsey-Fox method presented by Sitton, Burrus, Fox, and Treitel (2003). Computing the dominance of the resulting points with respect to all other curves is $d^2 t$ as well. Therefore the time complexity of Algorithm FindP is the complexity of Algorithm FindFunc, $\mathcal{O}((\frac{N}{k})^3)$, with additional cost of $\mathcal{O}(t^2 d^2) = \mathcal{O}((\frac{N}{k})^4)$ (the algorithm itself), i.e., jointly $\mathcal{O}((\frac{N}{k})^4)$.

**Theorem 8.** *Algorithm* FindP$(F, t)$ *finds point $p$ yielding the maximin value of* $\mathsf{ppd}$.

*Proof.* Algorithm FindP checks all intersection points between the pair of curves, and the points of local maxima of the curves. It then checks the dominance of these points, i.e., whether in the location these points have a lower value compared to all other curves, and picks the maximal of them. Therefore, if such a point is found, by Lemma 6, this point is precisely the maximin point. Moreover, by Lemma 7 this point exists. $\square$

### 4.4 Examples

We have fully implemented Algorithm FindP in order to find the optimal maximin $p$ for pairs of $d$'s and $t$'s. We use the following examples to illustrate how the relation between $t$ and $d$ is reflected in the $\mathsf{ppd}$ values. Recall that when running a deterministic patrol algorithm in

---

**Algorithm 2** Algorithm FindP($d, t$)

---
1: $F \leftarrow$ Algorithm FindFunc($d, t$).
2: Set $p_{opt} \leftarrow 0$.
3: **for** $F_{pivot} \leftarrow F_{1,\ldots,d}$ **do**
4:     Compute local maxima $(p_{max}, F_{pivot}(p_{max}))$ of $F_{pivot}$ in the range $(0,1)$.
5:     **for** each $F_i$, $1 \leq i \leq d$ **do**
6:         Compute intersection point $p_i$ of $F_i$ and $F_{pivot}$ in the range $(0,1)$.
7:         **if** $F_{pivot}(p_i) > F_{pivot}(p_{max})$ and $F_{pivot}(p_i) \leq F_k(p_i) \forall k$ **then**
8:             $p_{opt} \leftarrow p_i$.
9:         **if** $F_{pivot}(p_{max}) > F_{pivot}(p_i)$ and $F_{pivot}(p_i) \leq F_k(p_i) \forall k$ **then**
10:        $p_{opt} \leftarrow p_{max}$.
11: Return $(p_{max}, F_{pivot}(p_{max}))$.

---

all scenarios we handle, the minimal ppd is 0. We assume the robots are initially heading to the clockwise direction.

First of all, we have seen that the minimal ppd achieved after running FindP was always more than 0. As $t/d \to 1$, i.e., $t$ increases, then the value of the maximin ppd increases, and vice versa, i.e., as $t/d \to 1/2$, then the value of the maximin ppd decreases. This can be seen clearly in Figure 5. In this case, we have fixed the value of $t$ to 8 and checked the maximin ppd for $9 \leq d \leq 15$. When $t/d$ is close to 1 ($d = 9, t = 8$) the maximin ppd $= 0.423$, and the value decreases to 0.05 when $t/d$ is close to $1/2$ ($d = 15, t = 8$). Similar results are seen if we fix the value of $d$ and check for different values of $t$.

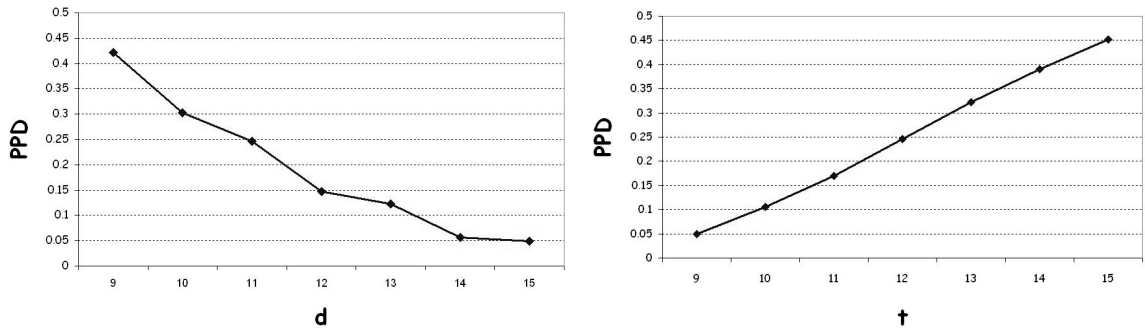

**Figure 5**: On the left, results of maximin ppd for fixed $t = 8$ and different values of $d$: the possible maximin ppd decreases as $d$ increases. On the right, results of maximin ppd for fixed $d = 16$ and different values of $t$: the possible maximin ppd increases as $t$ increases.

In Figure 6, we present the values of the ppd in all 16 segments, for all different possible values of $t$ ($9 \leq d \leq 15$). It is seen clearly, that the value of ppd usually decreases as the distance from the left robot increases, until it reaches the segment with maximin ppd, then the value rises again until reaching the current location of the robot to the right. The reason lies in the fact that the segments to the left of the segment with the maximin ppd are influenced mostly by the robot on the left, and the segments to the right of that point are mostly influenced by the robot to the right. Since the $p$'s yielding the maximin point in this example have value of greater than 0.8 for all $t$'s, the segment having the maximin value is to the right of the midpoint.

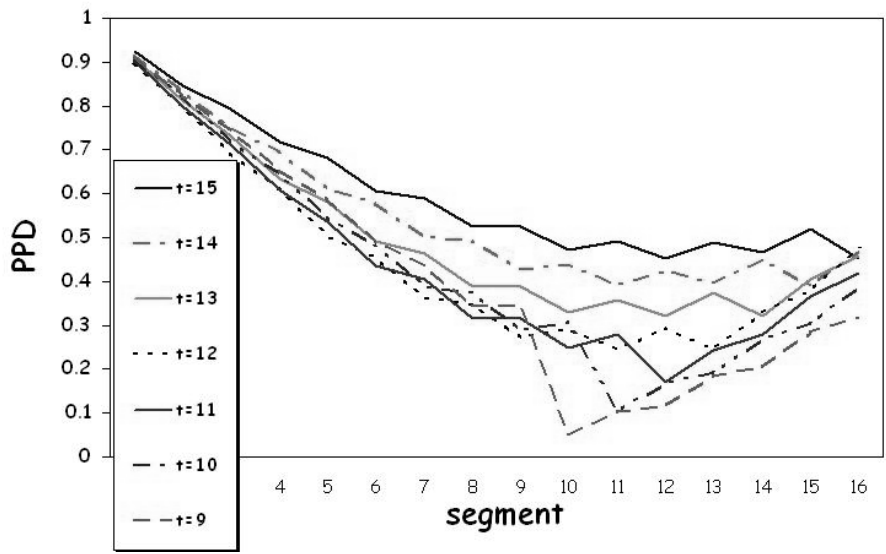

**Figure 6**: ppd values in all 16 segments for all $t$ values (9 to 15)

# 5. Accounting for Movement Constraints and Sensing Uncertainty

In this section we describe various ways in which the basic framework of multi-robot patrol can be used to solve the problem of finding an optimal patrol algorithm in various other settings. First, we describe the case in which the movement model of the robots is not necessarily directed. We then discuss various sensing capabilities of the robots in perimeter patrol: imperfect local sensing, perfect long-range sensing and imperfect long-range sensing. Finally, we describe the case in which the robots should travel along an open polyline (fence) rather than a perimeter.

## 5.1 Different Movement Models

A basic assumption of the robotic framework is that the robots' movement model is directed in the sense that if a robot has to go back to visit a point behind it, it has to physically turn around. This directed movement model is suitable for various robotic types, for example differential drive robots commonly used in robotic labs. However, in some cases the robots movement is undirected, for example if the robot travels along train tracks.We will demonstrate in this section how the basic framework can be used also in the latter case, i.e., if the robot movement is undirected.

We examine the difference in the Markov chain and the resulting ppd in three different cases:

1. Bidirectional Movement model, denoted by BMP. Here, the robots movement pattern is similar to movement on tracks or a camera going back and forth along a fixed course (omnidirectional robots). In this model, the robots have no movement directionality in the sense that switching directions—right to left and vice versa—does not require physically changing the direction of the robot (turning around).

2. Directional Costly-Turn model, denoted by DCP, the basic framework discussed to far for $\tau \geq 1$. The robots' movement is directed, and turning around is a special operation that has an attached cost in time. Specifically, we show the results here for $\tau = 1$.

3. Directional Zero-Cost model, denoted by DNCP, which is a special case of the DCP model with $\tau = 0$. The robots' movement is directed, yet turning around does not take extra time. This is coherently different from BMP, as in each step the robot does not go either right or left, but straight or back (where each could be either to the right or to the left, depending on the current heading of the robot).

The basic framework can be used for handling all three models simply by adapting the Markov chain to the current model. This changes only lines $5-6$ in Algorithm FindFunc. A description of the Markov chains are described in Figure 7. In the BMP model, it moves one step to the right (segment $i + 1$) with a probability of $p$ and one step to the left (segment $i-1$) with a probability of $1-p$. This model is similar to a random walk. The corresponding Markov chain is simple: edges exist from $s_i$ to $s_{i+1}$ with a probability of $p$ and from $s_i$ to $s_{i-1}$ with a probability of $1 - p$ (with no related direction). In both the DNCP and DCP models, we assume directionality of movement, hence the robot continues its movement in its current direction with a probability of $p$, and turns around (rewinds) with a probability of $1-p$. In DCP, if the robot turns around it will remain in segment $i$ (as described in Figure 2). In the DNCP model, the chain is similar to the one above, however edges will exist from $s_i^{cw}$ to $s_{i+1}^{cc}$ and from $s_i^{cc}$ to $s_{i-1}^{cw}$ with a probability of $1 - p$. See Figure 7 for an illustration of DNCP, DCP and BMP as a Markov chain.

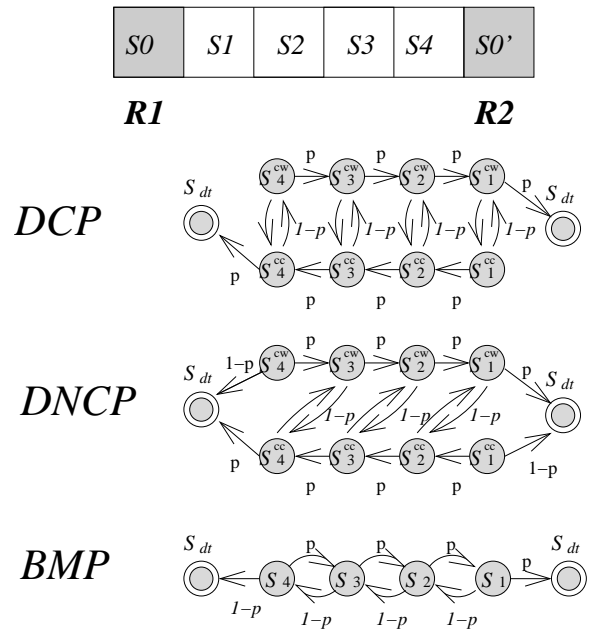

**Figure 7**: Conversion of the initial segments and robot locations into a graphical model in all three movement models.

We examined the difference between the resulting ppd values in the three models in a case where $d = 16, t = 12$ (Figure 8). It is clearly noticeable that the DCP model yields less or equal values of ppd compared to DNCP model throughout the segments. The reason is because when turning around, in the DCP model, the operation costs an extra cycle, therefore the probability of arriving at a segment decreases, compared to the case in which turning around is not costly. Another interesting phenomena is that the ppd values of the BMP are considerably higher (and close to 1) than the values obtained by other models for segments closer to the location of the righthand side robot. The value then decreases dramatically around the value of $t$ and then increases back again. Recall that here there is no directionality of movement, therefore the probability of going right is 0.707 and going left is $1 - 0.707 = 0.293$, which explains this phenomena. One might have expected to have $p = 0.5$ in the random walk model (BMP), however by choosing an equal probability for going right and left, the robots will necessarily neglect the segments further away from them (the mid segments between two consecutive robots), resulting in a lower minimal ppd.

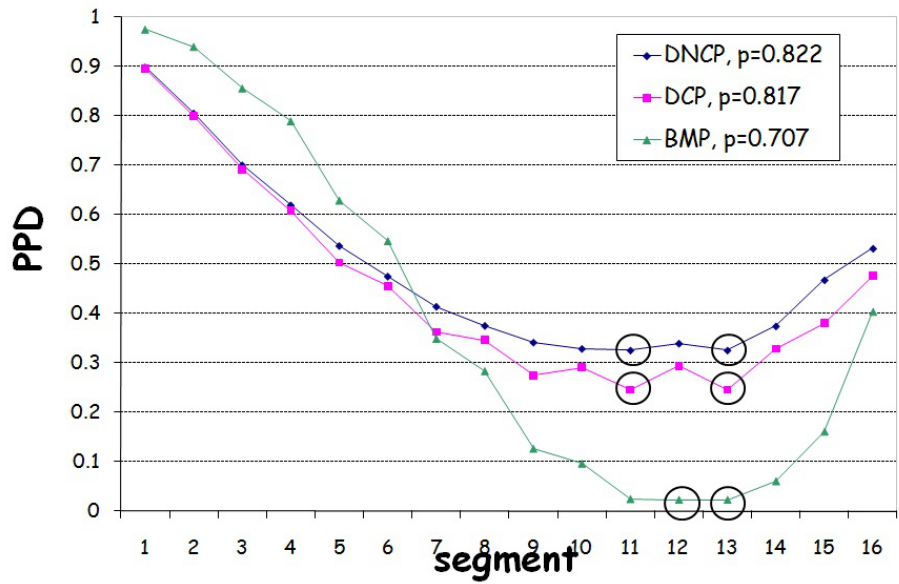

**Figure 8**: Results of maximin ppd values for $d = 16$ and $t = 12$ for all three models: DNCP, DCP and BMP. The maximin ppd values are circled.

## 5.2 Perimeter Patrol with Imperfect Penetration Detection

Uncertainty in the perception of the robots should be taken into consideration in practical multi-robot problems. Therefore we consider the realistic case in which the robots have imperfect sensorial capabilities. In other words, even if the adversary passes through the sensorial range of the robot, it still does not necessarily detect it.

We introduce the ImpDetect model, in which a robot travels through one segment per time cycle along the perimeter while monitoring it, and has imperfect sensing. Denote the probability that an adversary penetrating through a segment $s_i$ while it is monitored by some robot $R$ and $R$ will actually detect it by $p_d \leq 1$.

Note that if $p_d < 1$, revisiting a segment by a robot could be worthwhile—it could increase the probability of detecting the adversary. Therefore the probability of detection in a segment $s_i$ ($\mathsf{ppd}_i$) is *not* equivalent to the probability of *first* arriving at $s_i$ (as illustrated in Section 4.2), but the probability of detecting the adversary during *some* visit $y$ to $s_i$, $0 \leq y \leq t$. Denote the probability of the $y$'th visit of some robot to segment $s_i$ by $w_i^y$. Therefore $\mathsf{ppd}_i$ is defined as follows.

$$\mathsf{ppd}_i = w_i^1 p_d + w_i^1 (1 - p_d) \times \{w_i^2 p_d + w_i^2 (1 - p_d) \times \{\dots \{w_i^t \times p_d\}\}\} \tag{1}$$

In other words, the probability of detecting the penetration is the probability that it will be detected in the first visit ($w_i^1 \times p_d$) plus the probability that it will *not* be detected then, but during later stages. This again is the probability that it will be detected during the second visit ($w_i^2 \times p_d$) or at later stages, and so on.

Note that after $t$ time units, $w_i^t = 0$ for all currently unoccupied segments $s_i$, and if a robot resides in $s_i$, then $w_i^t$ is precisely $(1 - p_d)^t$.

One of the building blocks upon which the optimal patrol algorithms are based, is the assumption that the probability of detection decreases or remains the same as the distance from a robot increases, i.e., it is a monotonic decreasing function. This fact was used in Section 4 in proving that in order to maintain an optimal $\mathsf{ppd}$, the robots must be placed uniformly around the perimeter (with a uniform time distance), and maintain this distance by being coordinated. In order to show this here as well, we first prove that the probability of detection monotonically decreases with the distance from the location of the robot.

**Lemma 9.** *Let $S = \{s_{-t+\tau}, \dots, s_{-1}, s_0, s_1, \dots, s_t\}$ be a sequence of $2t$ segments, where robot $R_a$ resides in $s_0$ at time 0. Then $\forall i \geq 0$, $\mathsf{ppd}_i \geq \mathsf{ppd}_{i+1}$, and $\forall i \leq 0$, $\mathsf{ppd}_i \geq \mathsf{ppd}_{i-1}$.*

*Proof.* First, assume that $i > 0$ (positive indexes). By Equation 1, we need to compare between $w_i^1 p_d + w_i^1 (1 - p_d) \times \{w_i^2 p_d + w_i^2 (1 - p_d) \times \{\dots \{w_i^t \times p_d\}\}\}$ and $w_{i+1}^1 p_d + w_{i+1}^1 (1 - p_d) \times \{w_{i+1}^2 p_d + w_{i+1}^2 (1 - p_d) \times \{\dots \{w_{i+1}^t \times p_d\}\}\}$. It is therefore sufficient to show that $w_i^m \geq w_{i+1}^m$, for all $1 \leq m \leq t$. We prove this by induction on $m$. As the base case, consider $m = 1$, i.e., we need to show that $w_i^1 \geq w_{i+1}^1$. This is accurately proven in Lemma 1, based on the fact that the movement of the robots is continuous, therefore in order to get to a segment you must visit the segments in between (the formal proof also uses the conditional probability law).

We now assume correctness for $m' < m$, and prove that $w_i^m \geq w_{i+1}^m$. Denote the probability that a robot placed at segment $s_i$ will return to $s_i$ within $r$ time units by $x_i(r)$. In our symmetric environment, for every $i$ and $j$, $x_i(r) = x_j(r)$. Moreover, $\forall r, x_i(r) \geq x_i(r-1)$. Therefore $w_i^m$ can be described as $\sum_{r+u \leq t} w_i^{m-1}(u) \times x_i(r)$, and similarly $w_{i+1}^m = \sum_{r+u \leq t} w_{i+1}^{m-1}(u) \times x_{i+1}(r)$. By the induction assumption, $w_i^{m-1} \geq w_{i+1}^{m-1}$, and since $x_i(r) = x_{i+1}(r)$, it follows that $w_i^m \geq w_{i+1}^m$, proving the lemma for positive indexes.

The negative indexes are a reflective image of the positive indexes, but with $t - \tau$ time units. Since the induction was proven for all $t$ values, the proof for the negative indexes directly follows. □

The following Theorem follows directly from Lemma 9. The idea behind this is that since the probability of penetration detection decreases as the distance from the robots grow, both minimal ppd and average ppd are maximized if the distance between the robots is as small as possible. Since the patrol path is cyclic, this is achieved only if the distance between every two consecutive robots is uniform, and remains uniform. Note that Theorem 10 below is a generalization of Theorem 3 for imperfect sensing (based on the fact that that the general structure of the ppd function remains the same even if the robots might benefit from revisiting a segment, and by that increasing the ppd in that segment).

**Theorem 10.** *In the full knowledge adversarial model, a patrol algorithm in the* ImpDetect *model is optimal only if it satisfies two conditions: a. The robots are placed uniformly around the perimeter. b. The robots are coordinated in the sense that if they turn around, they do it simultaneously. By assuring these two conditions, the robots preserve a uniform distance between themselves throughout the execution.*

**Algorithm for finding ppd$_i$ with imperfect sensorial detection:**

Find the probability of penetration detection with $p_d \leq 1$ results in a different Markov chain, hence a different stochastic matrix $M$. Figure 9 demonstrates the new graphical model and the new resulting stochastic matrix $M$ (compared to Figure 2, in which $p_d = 1$). The difference in the algorithm is in the division of $s_0$ to two states, $s_0^{cw}$ and $s_0^{cc}$, the addition of the absorbing state $s_{dt}$ that represents the *detected* state and the transitions between these states. The ppd$_i$ is therefore obtained after $t+1$ steps (compared to $t$ steps) in the $s_{dt}$'s location in the result vector.

The time complexity of the algorithm remains $\mathcal{O}(d^4)$.

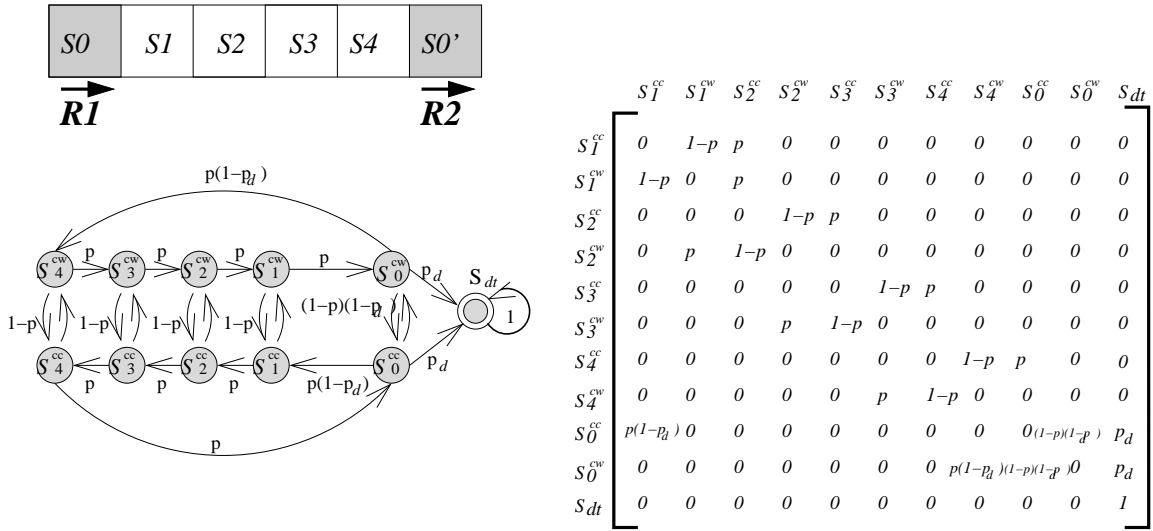

| | $s_1^{cc}$ | $s_1^{cw}$ | $s_2^{cc}$ | $s_2^{cw}$ | $s_3^{cc}$ | $s_3^{cw}$ | $s_4^{cc}$ | $s_4^{cw}$ | $s_0^{cc}$ | $s_0^{cw}$ | $s_{dt}$ |
|---|---|---|---|---|---|---|---|---|---|---|---|
| $s_1^{cc}$ | 0 | $1-p$ | $p$ | 0 | 0 | 0 | 0 | 0 | 0 | 0 | 0 |
| $s_1^{cw}$ | $1-p$ | 0 | $p$ | 0 | 0 | 0 | 0 | 0 | 0 | 0 | 0 |
| $s_2^{cc}$ | 0 | 0 | 0 | $1-p$ | $p$ | 0 | 0 | 0 | 0 | 0 | 0 |
| $s_2^{cw}$ | 0 | $p$ | $1-p$ | 0 | 0 | 0 | 0 | 0 | 0 | 0 | 0 |
| $s_3^{cc}$ | 0 | 0 | 0 | 0 | 0 | $1-p$ | $p$ | 0 | 0 | 0 | 0 |
| $s_3^{cw}$ | 0 | 0 | 0 | $p$ | $1-p$ | 0 | 0 | 0 | 0 | 0 | 0 |
| $s_4^{cc}$ | 0 | 0 | 0 | 0 | 0 | 0 | 0 | $1-p$ | $p$ | 0 | 0 |
| $s_4^{cw}$ | 0 | 0 | 0 | 0 | 0 | $p$ | $1-p$ | 0 | 0 | 0 | 0 |
| $s_0^{cc}$ | $p(1-p_d)$ | 0 | 0 | 0 | 0 | 0 | 0 | 0 | 0 | $(1-p)(1-p_d)$ | $p_d$ |
| $s_0^{cw}$ | 0 | 0 | 0 | 0 | 0 | 0 | 0 | $p(1-p_d)$ | $(1-p)(1-p_d)$ | 0 | $p_d$ |
| $s_{dt}$ | 0 | 0 | 0 | 0 | 0 | 0 | 0 | 0 | 0 | 0 | 1 |

**Figure 9**: Conversion of the initial segments and robot locations into a graphical model, and the respective stochastic matrix $M$ for the imperfect sensing model.

### 5.3 Improving Sensing Capabilities in Perimeter Patrol

In this section we present further enhancements by considering various sensing capabilities of the robots. Specifically, we first consider the case in which a robot can sense beyond its currently visited segment. We then offer a solution to the case in which the robot can sense beyond its current position, yet its sensing capabilities are not perfect, and change as a function of the distance from its current position.

#### 5.3.1 Extending (Perfect) Sensing Range

In this section we consider the LRange model, in which the sensorial range of a robot exceeds the section which it currently resides in. Use $L$ to denote the number of segments the robot senses beyond the segment it currently occupies. If $L > 0$, we refer to the $L$ segments as *shaded segments*. Note that the location of the shaded segments depends on the direction of the robot shading them, and they are always in the direction the robot is facing.

A trivial solution to dealing with such a situation is to enlarge the size of the segment, and thus enlarge the length of the time unit used as base for the system, such that it will force $L$ to be 0. However, in this case we lose accuracy of the analysis of the system, as the length of the time cycle should be as small as possible to also suit the velocity of the robots and the value of $t$.

In general, the values of $t$ that can be handled by the system are bounded by its relation to $d$ (the distance between every two robots along the path) - see Section 4. If $L > 0$, this changes. Specifically, if $L = 0$, then the possible values of $t$ considered are $\frac{d+\tau}{2} \leq t \leq d-1$. However, if $L > 0$, then it is possible to handle even smaller values of $t$, i.e., even if the penetration time of the adversary is short. Formally, the possible values of $t$ are given in the following equation.

$$\frac{d+\tau}{2} - L \leq t \leq d - L - 1$$

If $t$ is smaller than $\frac{d+\tau}{2} - L$, then an adversary with full knowledge will manage to penetrate with a probability of 1, i.e., there is a segment ($s_{L+1}$) which is unreachable within $t$ time units. On the other hand, if $t$ is greater than $d - L - 1$, then a simple deterministic patrol algorithm will detect all penetrations with a probability of 1. We assume that during the $\tau$ time units the robot turns around, it can sense only its current segment.

This change in the sensing model of the robot is reflected in the Markov chain, as seen in Figure 10. The change is that we add $2L$ arrows to the absorbing state $s_{dt}$, from $s_1^{cw}, \ldots, s_L^{cw}$ and $s_d^{cc}, \ldots, s_{d-L+1}^{cc}$. The stochastic matrix $M$ changes accordingly, and the probability of penetration detection in segment $s_i$ becomes the result of the vector multiplication $M^{t+1}V_i$, where $V_i$ is a vector of size $2d+1$ with all entries 0 except for entry corresponding to the location of $s_i^{cw}$, which holds a value of 1, similar to the process described in Algorithm FindFunc (1).

#### 5.3.2 Extending the Sensorial Range Along With Imperfect Detection

In many cases, the actual sensorial capabilities of the robot are composed of the two characteristics described in the previous sections, i.e., the robot can sense beyond its current segment, however the sensing ability is imperfect. Therefore in this section we introduce the ImpDetLRange sensorial model, which is a combination of the LRange and the ImpDetect models. Here the robot can sense $L$ segments beyond its current segment, yet the $p_d$ in each segment varies and is not necessarily 1. We therefore describe how to compute $\mathsf{ppd}_i$ in this case, which deals with the most realistic form of sensorial capabilities (Duarte & Hu, 2004): imperfect, long range sensing.

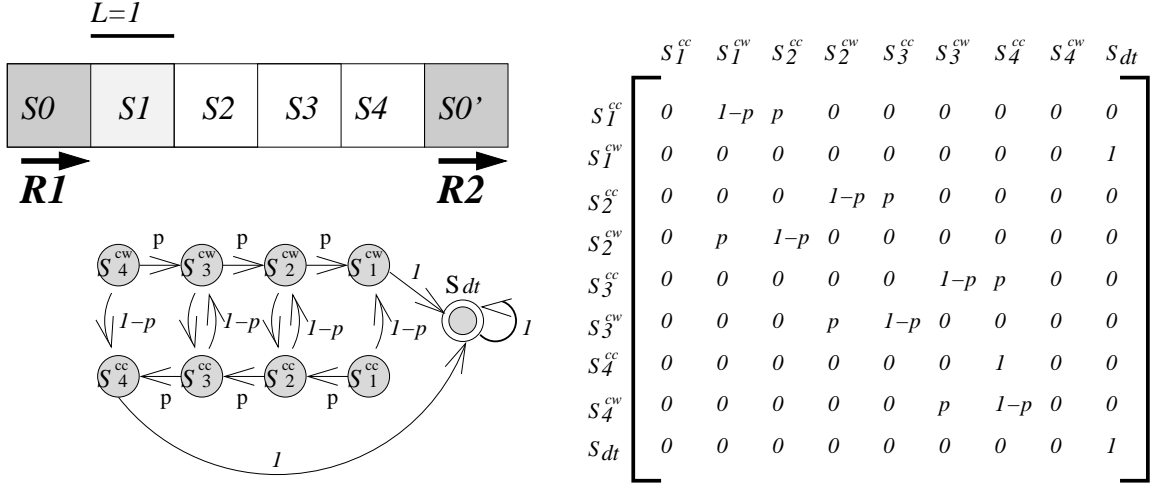

**Figure 10**: An illustration of $L$ segments shaded by robot $R$. In this case $R$ is facing right, therefore the shaded segments are to its right. The Markov chain changes accordingly, therefore also the stochastic matrix $M$.

The information regarding the sensorial capabilities of the robots includes two parameters. The first describes the quantity of the sensing ability, i.e., the number of segments that exceeds the current segment in which robot resides, for which it has *some* sensing abilities, denoted by $L$. The second parameter describes the quality of sensing in all segments the robot can sense. This is given in the form of a vector $V_S = \{v_0, v_1, \ldots, v_L\}$, where $v_i$ is the probability that the robot residing in $s_0$ will detect a penetration that occurs in segment $s_i$. We assume that the values in $V_S$ decrease monotonically, i.e., as $i$ increases, $v_i$ decreases or remains the same.

The Markov chain in this model, as illustrated in Figure 11, changes in order to reflect the imperfect sensing along with the long range sensing. The absorbing state $s_{dt}$ exist in addition to the states $s_0^{cw}$ and $s_0cc$. The transition probabilities are added from $2L$ segments: $\forall i, j \ 0 \leq i \leq L; d - L + 1 \leq j \leq d$, a transition from $s_i^{cw}$ to $s_0$ with probability $v_i$ and from $s_j^{cc}$ to $s_0$ with probability $v_{d-j+1}$. In addition, the transition from $s_i^{cw}$ to $s_icc$ is with probability $(1-p)(1-v_i)$, from $s_j^{cc}$ to $s_j^{cw}$ with probability $(1-p)(1-v_{d-j+1})$, hence the transition probability from $s_i^{cw}$ to $s_{i-1}^{cw}$ is $p(1-v_i)$ and from $s_j^{cc}$ to $s_{j+1}^{cc}$ is $p(1-v_{d-j+1})$.

The probability of penetration detection in segment $s_i$ is the result of $M^{t+1}$ multiplied by $V_i$ in location $s_0$ of the result vector. Note that also here, similar to the solution described in Section 5.2, since we added a new absorbing state (which takes an extra step to reach), $\mathsf{ppd}_i$ is the result in the product of the stochastic matrix and $V_i$ in location $s_0$ after $t + 1$ time steps (not $t$).

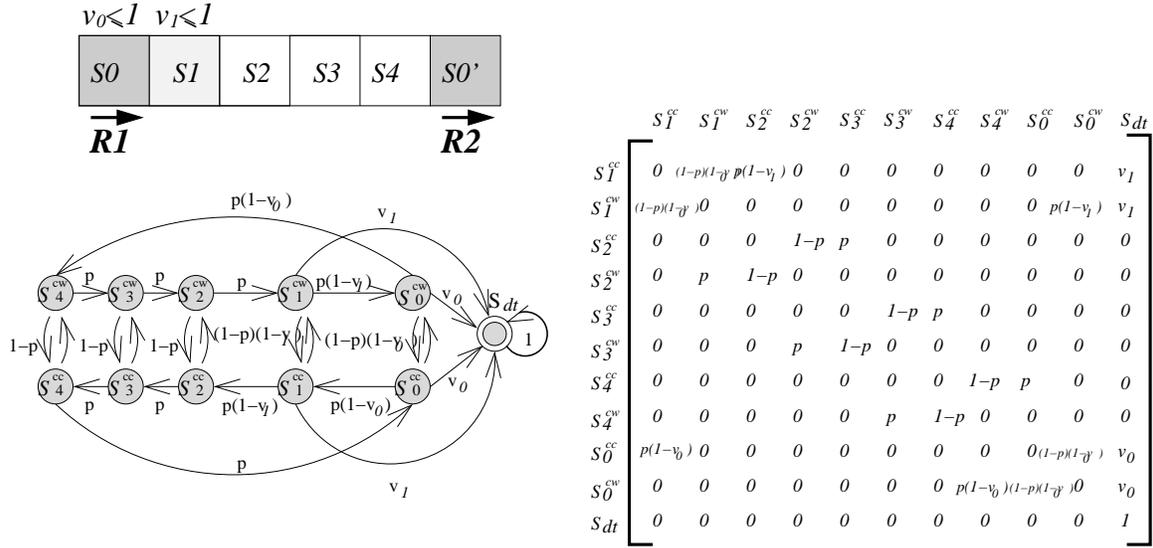

**Figure 11**: An illustration of $L$ segments shaded by robot $R$, where the probability of detection is not necessarily 1. In this case $R$ is facing right, therefore the shaded segments are to its right. The Markov chain and the stochastic matrix $M$ changes accordingly.

## 5.4 Multi-Robot Adversarial Patrolling Along Fences

In our general work, and specifically in previous sections, we assumed the robots travel around a closed, circular, area. In this section we discuss patrolling along an open polyline, also known as *fence patrol*. First, we will discuss how this patrol is different from perimeter patrol. We will then describe an algorithm for determining $\mathsf{ppd}_i$ in fence patrol assuming the robots have perfect sensing capabilities, and finally we will provide an algorithm for robots with imperfect sensing.

### 5.4.1 PATROLLING ALONG A CLOSED POLYLINE VS. AN OPEN POLYLINE

In the following, we describe why patrolling along an open polyline is more challenging than patrolling in cyclic environments (closed polyline).

The first reason lies in the fact that the robots are required to go back and forth along a part (or parts) of the open polyline. As a result, the elapsed time between two visits of a robot at each point along this line can be almost twice as long as the elapsed time in a circular setting. In Figure 12, we are given two environments: a closed polyline (circle) (a) and an open polyline (b). Note that open polylines b. and c. are equivalent in the sense that each robot travels through one segment per time step, regardless of the shape of the section. Both lines a. and b. are of the same total length $l$ and with the same number of robots (4). In the circular environment, if it takes an adversary more than $l/4$ time units to penetrate - it will never be able to penetrate even if the robots simply continuously travel with uniform distance between them. However, if the robots travel along an open polyline (b), the maximal time duration between two visits of the robot—even in the best case, is $2l/4 - 2$ (Elmaliach, Shiloni, & Kaminka, 2008). Therefore a weaker adversary that has a penetration time which is almost twice as long as in the circular fence might still be able to penetrate.

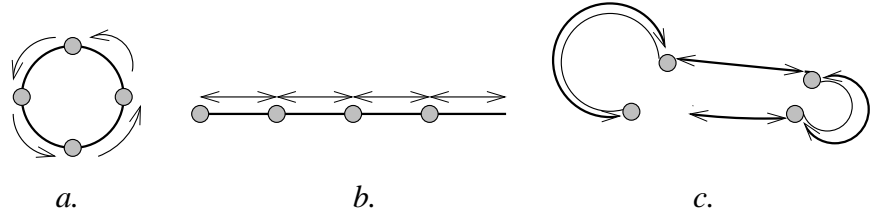

**Figure 12**: Illustration of the difference between patrolling along a line and patrolling along a circle, for different polylines

Another reason for the added complication in analyzing the probability of penetration detection in open polyline environments lies in the asymmetric nature of traveling in the segments along time. In a circular environment, if the robots are coordinated and switch directions in unison, then the placement of the robots is symmetric in each time unit. Therefore all segments in the same distance from some robot (with respect to its direction) have the same probability of penetration detection. Hence in order to calculate an optimal way of movement (in our case the probability $p$ of turning around), it is sufficient to consider only one section of $d$ segments, and the resulted $p$ is equivalent throughout the execution. In an open polyline environment this is not the case. The probability of penetration detection differs with respect to the current location and direction of the robot. Therefore the algorithm that finds the ppd for each segment, needs to calculate the ppd as a function of $p$ for each segment $s_i$ for each possible initial location of the robot inside the section. Therefore this results in a matrix of size $d \times d$ of the ppd functions (as opposed to a vector of $d$ functions in the circular fence).

### 5.4.2 DETERMINING $\mathsf{ppd}_i$ IN AN OPEN POLYLINE (FENCE)

Following the framework for multi-robot patrol along an open line proposed by Elmaliach et al. (2008), we assume each robot is assigned to patrol back and forth along one section of $d$ segments. Given this framework, we would like to compute the optimal patrol algorithm for the robots along the section. Similar to the perimeter patrol case (Section 4.2), we describe the system as a Markov chain (see Figure 13), with its relative stochastic matrix $M$. Since the robots have directionality associated with their movement, we create two states for each segment: the first for traveling in a segment in the clockwise direction, and the second for traveling in the counterclockwise direction. The probability of turning around at the end of each section is 1, otherwise the robot will continue straight with probability of $p$, and will turn around with probability of $1 - p$.

Note that the main difference from the perimeter patrol calculation of $\mathsf{ppd}_i$ lies in the number of resulting $\mathsf{ppd}_i$ functions. In perimeter patrol, due to its symmetric nature, there is one $\mathsf{ppd}_i$ function for each segment between the current location of each robot, representing the probability of a robot arriving there during $t$ time units. Here, however, $\mathsf{ppd}_i$ depends on the current location of the robot, hence for each location of the robot we have $d$ functions of probability of penetration detection, therefore a total of $d^2$ such functions (compared to $d$ in perimeter patrol).

Denote the probability of penetration detection in segment $s_i$ given that the robot is currently at segment $s_j$ by $\mathsf{ppd}_i^j$. In order to calculate the $d$ $\mathsf{ppd}_i^j$ function for all $1 \leq i, j \leq d$, we create $d$ different matrices: $M_1, \ldots, M_d$. Each matrix $M_i$ corresponds to calculating $\mathsf{ppd}_i^j$, i.e., the probability of penetration detection in segment $s_i$, and from that we calculate $\mathsf{ppd}_i^j$ from every current location $s_j$ of the robot (similar to what is done in perimeter patrol). Figure 13 demonstrates the matrix $M_2$ with which $\mathsf{ppd}_i^2$ is calculated. The figure describes the general case of $p_d \leq 1$, i.e., the robot might have imperfect sensing.

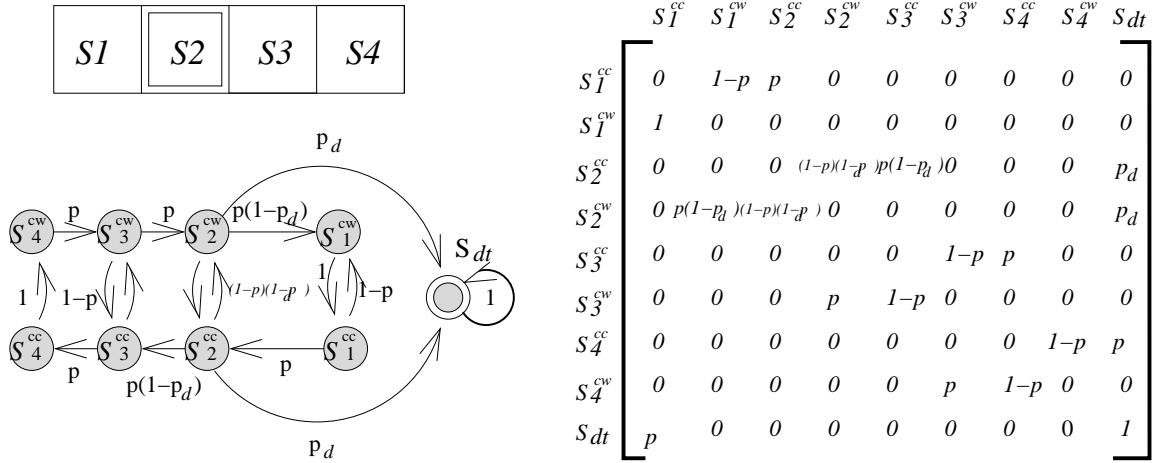

|  | $S_1^{cc}$ | $S_1^{cw}$ | $S_2^{cc}$ | $S_2^{cw}$ | $S_3^{cc}$ | $S_3^{cw}$ | $S_4^{cc}$ | $S_4^{cw}$ | $S_{dt}$ |
|---|---|---|---|---|---|---|---|---|---|
| $S_1^{cc}$ | $0$ | $1-p$ | $p$ | $0$ | $0$ | $0$ | $0$ | $0$ | $0$ |
| $S_1^{cw}$ | $1$ | $0$ | $0$ | $0$ | $0$ | $0$ | $0$ | $0$ | $0$ |
| $S_2^{cc}$ | $0$ | $0$ | $0$ | $(1-p)(1-p_d)$ | $p(1-p_d)$ | $0$ | $0$ | $0$ | $p_d$ |
| $S_2^{cw}$ | $0$ | $p(1-p_d)$ | $(1-p)(1-p_d)$ | $0$ | $0$ | $0$ | $0$ | $0$ | $p_d$ |
| $S_3^{cc}$ | $0$ | $0$ | $0$ | $0$ | $0$ | $1-p$ | $p$ | $0$ | $0$ |
| $S_3^{cw}$ | $0$ | $0$ | $0$ | $p$ | $1-p$ | $0$ | $0$ | $0$ | $0$ |
| $S_4^{cc}$ | $0$ | $0$ | $0$ | $0$ | $0$ | $0$ | $0$ | $1-p$ | $p$ |
| $S_4^{cw}$ | $0$ | $0$ | $0$ | $0$ | $0$ | $p$ | $1-p$ | $0$ | $0$ |
| $S_{dt}$ | $p$ | $0$ | $0$ | $0$ | $0$ | $0$ | $0$ | $0$ | $1$ |

**Figure 13**: Description of the system as a Markov chain, along with its stochastic matrix $M$ for calculating the ppd in segment $s_2$.

### 5.4.3 Optimal Algorithm for Fence Patrol

In the case of fence patrolling, the $\mathsf{ppd}$ value depends on the current location of the robot. Consequently, the optimal $p$ value characterizing the patrol of the robots is different for each segment $s_i$, where $1 \leq i \leq d$. Therefore there could be different optimal $p$ values with respect to both location and *orientation* of the robot ($2d$ values). However, it is sufficient to calculate the $\mathsf{ppd}$ values only $d$ times (and not $2d$ times)—only for one direction, as the other direction is a reflective image of the first.

In order to find the maximin point for the fence patrolling case, we use algorithm MaximinFence, which finds the value $p$ such that the minimal $\mathsf{ppd}$ is maximized, using Algorithm FindP that computes this point by finding the maximal point in the integral intersection of all curves ($\mathsf{ppd}_i$). The complete description of the algorithm is shown in Algorithm 3.

---

**Algorithm 3** Procedure MaximinFence($d, t$)

---

1: $M \leftarrow$ FindFencePPD($d, t$)
2: **for** $i \leftarrow 1$ to $d$ **do**
3:     $OpP[i] \leftarrow$ FindP($d, t$) with additional given input $M[i]$ as a vector of $\mathsf{ppd}$ functions.
4: Return $OpP$

---

## 6. Summary

This paper presents the problem of multi-robot patrolling in strong, full-knowledge, adversarial environments. In this problem a team of robots is required to repeatedly visit some path, in our basic case a perimeter, and detect penetrations that are controlled by an adversary. We assume the robots act in a strong adversarial model, in which the adversary has *full knowledge* of the patrolling robots and uses this knowledge in order to penetrate through the weakest spot of the patrol. We describe a framework for the basic case of multi-robot patrol around a closed polygon, and use this framework for developing, in polynomial time, an *optimal* patrol algorithm, i.e., an algorithm that strengthens the weakest spot of the patrol. This framework is then extended in order to solve the problem also in an open fence environment and in various movement and sensing models of the robots.

The paper makes several assumptions allowing the computation of an optimal strategy for the patrolling robots. One such assumption is the first order Markovian strategy of the patrolling robots. Although proving or disproving the optimality of using first order Markovian strategy is hard, it could be interesting to examine the case of higher order Markovian strategies and compare their time complexity and performance to the solution discussed here. Another direction for future work involves non-uniform environments, in which the utility obtained from detecting penetrations on one hand or succeeding in penetration on the other is not uniform throughout the environment. Other challenges left for future work include handling heterogenous robots and non linear environments.

## 7. Acknowledgments

Preliminary results appeared in Proceedings of the IEEE International Conference on Robotics and Automation (2008), in Proceedings of the Tenth Conference on Intelligent Autonomous Systems (2008) and in Proceedings of IJCAI Workshop on Quantitative Risk Analysis for Security Applications (QRASA) (2009). This research was supported in part by ISF grant #1357/07 and #1685/07, and MOST grant #3−6797. We thank the anonymous reviewers for constructive comments and helpful suggestions, and as always, thanks to K. Ushi.

## References

Agmon, N., Kaminka, G. A., & Kraus, S. (2008). Multi-robot fence patrol in adversarial domains. In *Proceedings of the Tenth Conference on Intelligent Autonomous Systems (IAS-10)*, pp. 193–201. IOS Press.

Agmon, N., Kraus, S., & Kaminka, G. A. (2008). Multi-robot perimeter patrol in adversarial settings. In *Proceedings of the IEEE International Conference on Robotics and Automation (ICRA)*.

Agmon, N., Kraus, S., & Kaminka, G. A. (2009). Uncertainties in adversarial patrol. In *Proc. of the IJCAI 2009 workshop on Quantitative Risk Analysis for Security Applications (QRASA)*.

Ahmadi, M., & Stone, P. (2006). A multi-robot system for continuous area sweeping tasks. In *Proceedings of the IEEE International Conference on Robotics and Automation (ICRA)*.

Almeida, A., Ramalho, G., Santana, H., Tedesco, P., Menezes, T., Corruble, V., & Chevaleyr, Y. (2004). Recent advances on multi-agent patrolling. *Lecture Notes in Computer Science, 3171*, 474–483.

Amigoni, F., Gatti, N., & Ippedico, A. (2008). Multiagent technology solutions for planning in ambient intelligence. In *Proceedings of Agent Intelligent Technologies (IAT-08)*.

Basilico, N., Gatti, N., & Amigoni, F. (2009a). Extending algorithms for mobile robot patrolling in the presence of adversaries to more realistic settings. In *Proceedings of the IEEE/WIC/ACM International Conference on Intelligent Agent Technology (IAT)*, pp. 565–572.

Basilico, N., Gatti, N., & Amigoni, F. (2009b). Leader-follower strategies for robotic patrolling in environments with arbitrary topologies. In *AAMAS*, pp. 57–64.

Chevaleyre, Y. (2004). Theoretical analysis of the multi-agent patrolling problem. In *Proceedings of Agent Intelligent Technologies (IAT-04)*.

Coppersmith, D., Doyle, P., Raghavan, P., & Snir, M. (1993). Random walks on weighted graphs and applications to on-line algorithms. *Journal of the ACM, 40*(3).

Duarte, M. F., & Hu, Y. H. (2004). Distance-based decision fusion in a distributed wireless sensor network. *Telecommunication Systems, 26*(2-4), 339–350.

Elmaliach, Y., Agmon, N., & Kaminka, G. A. (2007). Multi-robot area patrol under frequency constraints. In *Proceedings of the IEEE International Conference on Robotics and Automation (ICRA)*.

Elmaliach, Y., Agmon, N., & Kaminka, G. A. (2009). Multi-robot area patrol under frequency constraints. *Annals of Math and Artificial Intelligence, 57*(3-4), 293–320.

Elmaliach, Y., Shiloni, A., & Kaminka, G. A. (2008). A realistic model of frequency-based multi-robot fence patrolling. In *Proceedings of the Seventh International Joint Conference on Autonomous Agents and Multi-Agent Systems (AAMAS-08)*.

Gustavson, F. G. (1978). Two fast algorithms for sparse matrices: Multiplication and permuted transposition. *ACM Trans. Math. Softw., 4*, 250–269.

Haynes, T., & Sen, S. (1995). Evolving behavioral strategies predators and prey. In *IJCAI-95 Workshop on Adaptation and Learning in Multiagent Systems*, pp. 32–37.

Lynch, N. A. (1996). *Distributed Algorithms*. Morgan Kaufmann.

Paruchuri, P., Pearce, J. P., Tambe, M., Ordonez, F., & Kraus, S. (2007). An efficient heuristic approach for security against multiple adversaries. In *Proceedings of the Sixth International Joint Conference on Autonomous Agents and Multi-Agent Systems (AAMAS-08)*.

Paruchuri, P., Tambe, M., Ordonez, F., & Kraus, S. (2007). Security in multiagent systems by policy randomization. In *Proceedings of the Sixth International Joint Conference on Autonomous Agents and Multi-Agent Systems (AAMAS-07)*.

Pita, J., Jain, M., Ordonez, F., Tambe, M., Kraus, S., & Magorii-Cohen, R. (2009). Effective solutions for real-world stackelberg games: When agents must deal with human uncertainties. In *Proceedings of the Eighth International Conference on Autonomous Agents and Multiagent Systems (AAMAS-09)*.

Sak, T., Wainer, J., & Goldenstein, S. K. (2008). Probabilistic multiagent patrolling. In *Proc. of the 19th Brazilian Symposium on Artificial Intelligence (SBIA-08)*, pp. 124–133.

Sharir, M., & Agarwal, P. K. (1996). *Davenport-Schinzel sequences and their geometric applications*. Cambridge University Press.

Shieh, J. S., & Calvert, T. W. (1992). View and route planning for patrol and exploring robots. *Advanced Robotics, 6*(4), 399–430.

Sitton, G., Burrus, C., Fox, J., & Treitel, S. (2003). Factoring very-high-degree polynomials. *Signal Processing Magazine, IEEE, 20*(6), 27 – 42.

Stewart, W. J. (1994). *Introduction to the Numerical Solution of Markov Chains*. Princeton University Press.

Vidal, R., Shakernia, O., Kim, H. J., Shim, D. H., & Sastry, S. (2002). Probabilistic pursuit-evasion games - theory, implementation, and experimental evaluation. *Robotics and Automation, IEEE Transactions on, 18*(5), 662–669.